\documentclass[prc,showpacs,twocolumn,floatfix]{revtex4}
\usepackage{graphicx}
\usepackage{dcolumn}
\usepackage{bm}
\usepackage{multirow}
\usepackage{color}
\usepackage{amssymb}
\usepackage{natbib}
\begin{document}

\title{The size of most massive neutron stars may reveal its exotic cores}

\author{Neha Gupta and P. Arumugam}

\address{Department of Physics, Indian Institute of Technology Roorkee, Roorkee - 247 667, India}
\begin{abstract} 
The recent high precision observation of the most massive pulsar J1614-2230 with $(1.97\pm0.04)$ solar mass ($M_\odot$) was reported with a suggestion that many nuclear models which consider exotic particles in the core could be ruled out. However, many recent calculations could explain this star with various exotic particles, rendering the precise mass measurements insufficient to conclude on exotic cores. We examine the sensitivity of the radius of such a star to the details of its core.  With our calculations and analysis, here we show that, for the most massive neutron star, with a precise observation of its radius it is possible to ascertain the presence of exotic cores.
\end{abstract}


\maketitle

The constituents of neutron star (NS), especially at its core remain as an intriguing subject for several decades. NS represents one of the final stage of the stellar evolution,
and considered as a laboratory for the study of  matter at super high densities.  Observing the thermal spectra of NS results in a measurement of the gravitational redshift, which depends on both radius and mass of the star. With one
observation the mass and radius can not be determined accurately~\cite{lattimer:426}. However, precise mass measurements of NS have been possible~\cite{demorest},
and significant work is in progress for   measuring both mass and radius with
more accuracy~\cite{Ozel:5,Steiner:33,Guillot:88,expt_m5,expt_m4}. Such measurements are crucial in improving our understanding of NS matter and the underlying interactions, because the observed masses and radii of NS constrain the equation of state (EoS) of NS or in general the nuclear matter
and the finite nuclear properties such as neutron skin thickness in heavy nuclei \cite{FSU_para}, breathing mode resonance energies
\cite{FSU_para,Fattoyev:055803} etc. 

In an era where the nuclear models and their parameters are expected to account for the properties of infinite matter and neutron star \cite{FSU_para,Fattoyev:055803,Shen:065808,furnstahl,nl3_para,ermf}, the new observation \cite{demorest} of the
pulsar J1614-2230 with mass $(1.97\pm0.04)\ M_\odot$ stands as a stringent test. This observation has a strong impact on
the neutron star (NS) related studies mainly due to the ruling out
of: (i) many of the standard and successful  EoS by then (for e.g. PAL6 \cite{Prakash:2518},
FSUGold \cite{FSU_para}) which were yielding softer EoS and hence lower mass, and (ii) most of the EoS involving exotic matter such as kaon condensates or hyperons (for e.g.~GM3 \cite{Broderick:2002167}, GS1 \cite{Schaffner:309}) which were ``believed"
to be in the core of NS. 
Soon, these claims were played
down by  J.M. Lattimer \cite{Naturenews} by stating \textit{``To rule out these exotic models fully you also need to know the radius of the star"}. He also added that it will be easy to tweak the parameters, to bring back the exotic particles and/or to reproduce the mass. This tweaking cannot be
arbitrary due to several established constraints for the symmetry energy,
EoS and finite nuclear
properties.  Many of the recent work were
in this direction where few more EoS were introduced \cite{Hempel:70}
and in several models
the parameters were adjusted \cite{Fattoyev:055803,Shen:065808}, to  reproduce the $2 M_\odot$  NS. However, in the context of confirming/ruling-out exotic cores in
neutron stars, the precision required in observation of radius of most
massive neutron stars, is not reported so far.

 In our recent work \cite{neha_antikaon} we found that for a NS with maximum
   mass $2 M_\odot$, without antikaons (G2) we get the radius is 11.03 km
whereas, with $K^-$
(FSU2.1)   and both antikaons
(G1) the radii are 11.42 km and 12.55 km, respectively. This sensitivity of the radius  to the presence of exotic
cores motivated us for a theoretical survey which may demand more precise observation  of radius along with the mass of such a heavy NS. In the following
text we outline the results from (i) our calculations with
recent versions of
extended relativistic mean field (RMF) models which are one among the most
reliable models for NS EoS, (ii) variety of interactions in literature and (iii) a fiducial model with parabolic EoS, all successfully explaining a static $2 M_\odot$  NS. In this work we have assumed that the maximum in observed
neutron star masses $[(1.97\pm0.04)M_\odot]$ to be same as the maximum mass allowed by a chosen EoS. It is likely that latter can be at least a few tenths
of solar mass larger \cite{Lattimer:preprint}.  Implications of this on our
results are studied with the fiducial model. 
\begin{figure*}
\centering
\includegraphics[width=.99\textwidth]{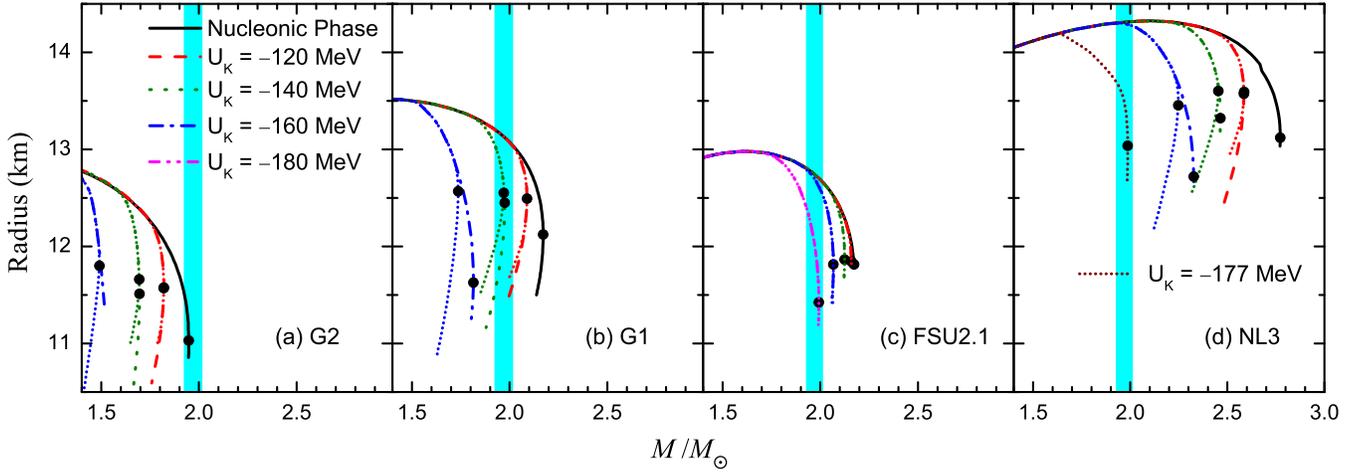}
\caption{The mass-radius relation from RMF models \cite{neha_antikaon}. Different curves represent the   calculations using different kaon optical potentials ($U_K$) and with different parameter sets. For each parameter set, solid black line represents nucleonic (non-kaonic) phase, lines with different
patterns and colors represent the phase with $K^-$ and the corresponding
small dotted lines represent the phase with both antikaons $(K^-,\bar{K^0})$. The different patterns and colors represent the strength of the kaon optical potential $U_K$ ($|U_K|$ quantifies the influence of kaons) as specified in the inset. The sensitivity of mass-radius relation to
the parameter $U_K$ depends on the stiffness of symmetry energy.  For e.g.,
FSU2.1 has softer symmetry energy and hence it is not much sensitive to $U_K$.
The onset of kaon condensation is governed by the interplay between density
dependance of, symmetry energy and EoS~\cite{Gupta:015804}. The delayed onset
of $K^-$ in FSU2.1, due to the above reason, forbids the onset of $\bar{K}^0$.
The solid circles represent the maximum mass in every case. Mass is given in units of solar mass ($M_\odot$). The shaded region correspond
to the recent observation of $(1.97\pm 0.04) M_\odot$ neutron star.}
\label{fig:mass}
\end{figure*}

\begin{table*}
\centering
\caption{Radii of  neutron
star of maximum mass $(1.97\pm0.04) M_\odot$ resulting from different models/parameters
and with different compositions as reported in the recent literature.}
\label{tab:observed_radius}
\begin{tabular}{lcc}\hline \hline
Model/Parameter & Radius (km) & Composition  \\\hline
APR (NS) \cite{Alford:024007}& 11.0 & $n, p, e^-, \mu^-$ \\
G2 \cite{Gupta:015804}  & 11.03 & $n, p, e^-, \mu^-$ \\
IU-FSU \cite{Fattoyev:055803} & 11.2 & $n, p, e^-, \mu^-$ \\
FSU2.1 \cite{neha_antikaon} & 11.42 &$n, p, e^-, \mu^-,K^-$ \\
G1 \cite{Gupta:015804} & 12.45 & $n, p, e^-, \mu^-,K^-$\\
G1 \cite{neha_antikaon} & 12.55 & $n, p, e^-, \mu^-,K^-,\bar{K^0}$\\
NL3 (This work) & 13.04 & $n, p, e^-, \mu^-,K^-,\bar{K^0}$ \\
Glendnh3+$\Gamma_{th}$ \cite{Bauswein:011101} & 11.48 & $n, p, e^-, \mu^-$+Hyperons \\
DD2H\cite{Lastowiecki:2011} & 11.6 & $n, p, e^-, \mu^-$+Hyperons \\
QMC700/NY\cite{Massot:39002}& 12.5 & $n, p, e^-, \mu^-$+Hyperons\\
QMC-HF NY$\kappa_I$-$\pi$-$m_{\sigma}^*$ \cite{Whittenbury}& 12.38& $n, p, e^-, \mu^-,\pi$+Hyperons\\
QCD ($a_4$=.62 and $\eta=-$1)\cite{Kim:035810}&10.60&$n, p, e^-, \mu^-,K^-$+
Quarks\\
APR (HS) \cite{Alford:024007}& 12.2 & $n, p, e^-, \mu^-$ + Quarks \\
%$SU(3)$ parity model\cite{Schramm:5113} & 10.4 & $n, p, e^-, \mu^-$+Hyperon+Quark\\
%GM1+ NJL\cite{Schramm:5113} & 10.4 & $n, p, e^-, \mu^-$+Hyperon+Quark\\
DD2H-NJL, set A\cite{Lastowiecki:2011} & 11.8 & $n, p, e^-, \mu^-$+Hyperons+ Quarks \\
DD2H-NJL, set B\cite{Lastowiecki:2011} & 12.15 & $n, p, e^-, \mu^-$+Hyperons+ Quarks \\
GM1+PNJL\cite{Shao:343} &13.2&$n, p, e^-, \mu^-$+Hyperons + Quarks\\\hline \hline
\end{tabular}
\end{table*}

In the RMF model, we have chosen  few representations, which
signify as extrema with and without antikaons ($K^-,\bar{K^0}$), that can yield $2 M_\odot$ as the maximum mass of NS.  The model Lagrangian
and details of calculation are explicitly explained elsewhere \cite{furnstahl,Gupta:015804,ermf}.   Using the Lagrangian density, we obtain the energy density and pressure (EoS) with or without antikaons. Once the EoS is defined, we  use the Tolman-Oppenheimer-Volkoff   equations \cite{tov1,tov} to get the mass-radius relation of a NS.

 The mass-radius relation, thus obtained,  using the parameters
NL3\cite{nl3_para},
G1, G2 \cite{furnstahl}, and FSU2.1\cite{Shen:065808} are shown in Fig.~1.
  Among these chosen parameters, G2 yields the softest EoS.  In general, inclusion of exotic particles soften
the EoS and a softer
EoS results in lesser mass for NS.  Without inclusion of any exotic particle, G2 yields a mass consistent
with the $2 M_\odot$ NS. Thus G2 represents one extremum where exotic particles cannot be accommodated.  The other extremum
should correspond to an EoS with reasonably extreme value of $U_K$ (if we
consider only antikaons as the exotic particles) yielding
a $2 M_\odot$ NS. NL3 represents this extremum and its EoS is the stiffest among the chosen parameters.  The other parameters G1 and FSU2.1
fall in between the extrema and have similar EoS but significantly different symmetry energy contribution. From Fig.~1,
%\ref{fig:mass}
one can observe a
general trend that the radius ($R$) corresponding to maximum mass is lower for a softer EoS. However, this scenario can change in the presence of antikaons.
 Though the antikaons soften the EoS, they can increase or decrease the central baryon density ($\rho_c$) \cite{Gupta:015804} and hence  $R$ ($R\propto 1/\rho_c$).
This results in, different patterns of variation of $R$, as shown in Fig.~1.

These results suggest that a $2M_\odot$ NS
can be explained with different quantity of antikaons, and the
radius $(R_2)$ corresponding to a NS with a maximum mass of $2M_\odot$
is different with different parameter sets. Our results from RMF models show that if $R_2 \gtrsim 11.04$ km then the exotic core must be present.  In our calculations exotic core is restricted to antikaons, but these calculations should be extended to
include hyperons and quarks.  Similar calculations for the $2M_\odot$ NS
with other models, and a careful scrutiny of the effect of rotation on massive
NS, are needed to have a complete outlook.

\begin{figure*}
\centering
\includegraphics[width=.99\textwidth]{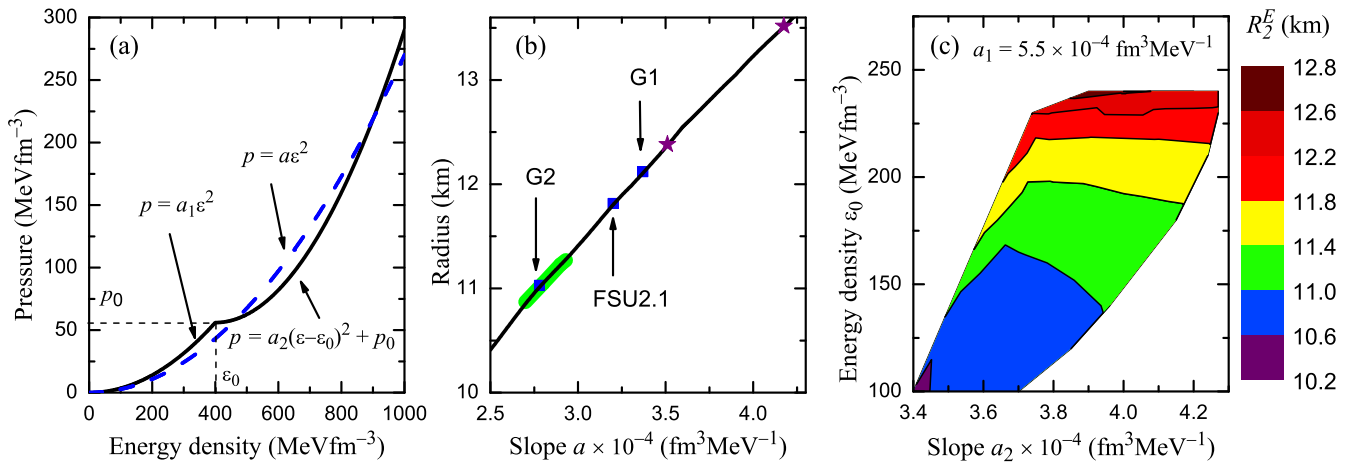}
\caption{The mass-radius relation from fiducial model. \textbf{(a):} Schematic representation of EoS from our fiducial model comprising one or two parabolas represented by dashed and solid lines respectively. The dashed
line does not change abruptly as a function of density.  This could effectively
mimic an EoS of NS with nucleonic phase only, with a single parameter ($a$) representing
the stiffness of the EoS. In case of EoS with two parabolas with slopes $a_1$
and $a_2$,
the point of intersection ($\varepsilon_0,p_0$) represents the onset of exotic phase whence the EoS changes abruptly. Both these EoS can yield same mass but different radii for a neutron star.
\textbf{(b):} The radius of the neutron star  (when its mass is maximum) plotted as a function of the slope of parabola when the EoS is assumed to be a single parabola. Green region is where the EoS yield a neutron star
with maximum mass $(1.97\pm0.04) M_\odot$, and blue squares indicate few selected RMF
models as shown in the labels. The first and second stars represent the cases
where the maximum masses are 2.2$M_\odot$ and 2.4$M_\odot$, respectively. \textbf{(c):} Sample results for the radius of a neutron star with maximum mass $(1.97\pm0.04) M_\odot$ obtained from an EoS with two parabolas, plotted as a function of the
slope of second parabola ($a_2$) representing the exotic phase and the point
at which the second parabola starts to contribute ($\varepsilon_0$).}
\label{fig:toy}
\end{figure*}

In Table \ref{tab:observed_radius}, we have compiled $R_2$ calculated with different interactions and compositions as quoted in recent literature.  We have excluded the cases which
yield the maximum mass higher than $2.01M_\odot$.  For further
discussions, it is convenient to define two radii $R_2^N$ and $R_2^E$ (in
km) for the cases of NS without exotic core and with exotic core respectively, corresponding to a maximum mass of $(1.97\pm 0.04) M_\odot$.  From Table~\ref{tab:observed_radius}
we have $10.6 \leq R_2^E \leq 13.2$ and $11.0 \leq
R_2^N\leq 11.2$.  Thus we can claim that if $R_2$ were observed beyond the
range of $R_2^N$ then the exotic core must be present. Otherwise, the existence of exotic
cores  will remain ambiguous. To
strengthen this claim, one has to look more into the systematics, beyond relying
on few theoretical results available till date.

As a first step, we start with a fiducial model where the EoS is a single
parabola represented by the dashed line
in Fig 2a.  The resulting radii are shown in Fig.~2b
%\ref{fig:a}
  as a function of slope of the parabola $(a)$.  Smaller the value $a$, softer the EoS, more
is the central density and hence we obtain a smaller NS. By varying $a$, we could
fit EoS of different RMF models. The radii obtained with such fitted EoS
are quite consistent with those obtained from the actual EoS, justifying
our approach.  These radii are marked in Fig~2b.
%\ref{fig:a} 
For a NS with mass $(1.97\pm0.04) M_\odot$, our fiducial model calculations predict that $2.7 \leq a \leq 2.93$ and correspondingly $10.87 \leq R_2^N \leq 11.27$.  However, this range is very inclusive because many values of
the parameter $a$ yield quite unrealistic EoS which will not satisfy several experimental
constraints.

In our next step we consider two parabolas for the EoS represented by the
solid line
in Fig 2a. Sample results for
  $R_2^E$ with a fixed slope of the first parabola ($a_1$) are given as a function of slope of the second parabola ($a_{2}$), and the energy density ($\varepsilon_0$) at which the second parabola starts to contribute to the EoS. We have considered a broader range of $\varepsilon_0$ whereas the lower limit of $\varepsilon_0$ shall correspond to quarks which typically start to contribute from $\sim 100$ MeVfm$^{-3}$. %\ref{fig:b}
 $\varepsilon_0$
and $a_2$ varied in tandem can effectively represent  the parameters of exotic
matter EoS.  In case of antikaons we found that they simulate the variation in $U_K$.
The important feature of Fig.~2c
%\ref{fig:b}
 is that the radius is quite sensitive
to the parameters of the exotic matter EoS and 
%\ref{fig:b}
 the variation
in $\varepsilon_0$
and $a_2$ leads to a change in radius up to $\sim 2.5$ km.  We can see that for a given $a_1$ and $a_2$, $R_2^E$ increases
with $\varepsilon_0$. Lower values of  $R_2^E$ are due to the early onset of exotic phase.   The range of $R_2^E$ obtained for the chosen $a_1$ value
is presented in Fig.~2c
%\ref{fig:b}
 but a complete range for all allowed values
of $a_1$, $a_2$ and $\varepsilon_0$ is given in Table~\ref{tab:radius}, where we summarize all of our results.

\begin{table}
\centering
\caption{Radius of a neutron star of mass $(1.97\pm0.04) M_\odot$, obtained from different models assuming this mass as the maximum mass.}
\label{tab:radius}
\begin{tabular}{lcc}\hline \hline
{Model} & \multicolumn{2}{c} {Radius (km)} \\ 
\cline{2-3} & Without exotic core & With exotic core \\\hline
RMF (Fig.~1)
%\ref{fig:mass}
& $ 11.03$ & $11.43 \leq R \leq 13.04$ \\
Parabolic EoS & $10.87 \leq R \leq 11.27$ & $9.79 \leq R \leq 13.22$ \\
%%X7 ($r_{ph}\gg R$)($M$=(1.62,2.08)$M_\odot$)& \multicolumn{2}{|c|} {$10.27 < R < 12.25$} \\\hline
From Table \ref{tab:observed_radius} & $11.0 \leq R \leq 11.2$ & $10.6 \leq R \leq 13.2$ \\\hline \hline
\end{tabular}
\end{table}

Our fiducial model suggests  that
without exotic core we have a narrow allowed range of $R_2$.  If $R_2$ were
observed beyond this range, for e.g. $R_2=12.3\pm1$ km or $10.3\pm0.5$ km,
then an exotic core must be present. The upper limit of this allowed range could be higher, if (i) a NS is observed with mass
above $1.97\pm.04M_\odot$ or (ii)  we consider that the maximum mass given by the EoS has to be higher than the maximum in observed
mass. Any increase
in mass will result in the increase of radius without the exotic cores [See
Fig.~2b].
 Consequently, cases
(i) and (ii) will have no bearing in the lower limit of the range, i.e.,
the least radius that could be supported by a NS without exotic cores.
 Hence, it is clear that if $R\lesssim 10.87$ km then the NS must have an exotic
 core. The range of radii obtained by the fiducial model may vary, if one considers
two polytropes instead of two parabolas but with two more parameters to adjust.
However, from Table~\ref{tab:radius} we clearly see that the ranges of $R_2$
 given by our fiducial model comprise the ranges obtained from our RMF calculations
 and other recent results.  Hence safer conclusions could be drawn from these
broader ranges of $R_2$.  If we consider more realistic form for the EoS and incorporate the experimental constrains, for
the saturation properties defining the EoS and the high density behaviour of the EoS (For e.g. \cite{expt_data}), the range of $R_2^N$ may become narrow.  
 
For a static neutron star with maximum  mass $(1.97\pm 0.04) M_\odot$, we conclude that a pure nucleonic star can have only
a narrow range of radius (For e.g. $10.8$ km $\lesssim R \lesssim\ 11.3$ km) and the presence of exotic cores widen this range.  Beyond the above-mentioned  range, exotic cores must be present. If the maximum mass given by an EoS
 has to be larger than the maximum in observed NS masses then the upper limit
of above range could be different but the constraint on the lower
limit remains intact.  More rigorous calculations are needed to obtain precisely
the range of radius allowed for the most massive neutron star without
exotic particles. 
 
The $2M_\odot$ NS does not rule out exotic cores and hence  the constituents
of NS still remain a puzzle. To overcome this, more observations and calculations should focus on the radius of such massive neutron stars. A precise measurement of mass and radius of a lower mass neutron star
will not be useful confirm the presence of exotic cores. Our calculations and analysis predict that a precise observation of radius
of a $2M_\odot$ NS is capable of confirming the presence of exotic cores.  This in turn can shed light on the high density behaviour
of matter and the phase transitions that can happen at such extreme environments.

\providecommand{\newblock}{}

\end{document}